\documentclass[final,3p,times]{elsarticle}

\usepackage{amssymb}
\usepackage{amsmath}
\usepackage{braket}
\usepackage[symbol]{footmisc}
\usepackage{wrapfig}

\def\dd{\mbox{d}}
\def\ddd{\textrm{d}}
\def\hh{\hspace{0.5mm}}

\journal{Physica A}

\begin{document}
\onecolumn
\begin{frontmatter}



\title{Doi-Peliti Path Integral Methods for Stochastic Systems with Partial Exclusion}

\author[1]{Chris D. Greenman\footnote{Corresponding author E-mail address: C.Greenman@uea.ac.uk}}
\address[1]{School of Computing Sciences, University of East Anglia, NR4 7TJ, United Kingdom.}

\begin{abstract}
Doi-Peliti methods are developed for stochastic models with finite maximum occupation numbers per site. We provide a generalized framework for the different Fock spaces reported in the literature. Paragrassmannian techniques are then utilized to construct path integral formulations of factorial moments. We show that for many models of interest, a Magnus expansion is required to construct a suitable action, meaning actions containing a finite number of terms are not always feasible. However, for such systems, perturbative techniques are still viable, and for some examples, including carrying capacity population dynamics, and diffusion with partial exclusion, the expansions are exactly summable.
\end{abstract}


\end{frontmatter}


\section{Introduction}
\label{Intro}

This work is concerned with parallels between quantum field theory (QFT) and population dynamics. QFT was developed  \cite{Weinberg1995}, \cite{Mandl2010} to model interactions of subatomic particles. These interactions result in particle populations that vary in size and position. Classical population dynamics also model populations that vary in size, via mechanisms such as birth-death processes, for example. These populations can also vary in `position', where position can be interpreted as a continuous feature of interest, such as physical location of a molecule, the size of a cell, or the age of individuals, for example. Doi \cite{Doi1976}, \cite{Doi1976b} was the first to notice this parallel and used QFT machinery to model molecular reactions. 

The path integral formulation of quantum mechanics was introduced by Dirac, further developed and popularized by Feynman \cite{Feynman1965}. Peliti \cite{Peliti1985} adapted these ideas, using functional integration techniques to construct path integral formulations of the Doi paradigm. These techniques have seen a range of applications including molecular reactions \cite {Doi1976}, \cite{Doi1976b}, birth-death processes on lattices \cite{Peliti1985}, \cite{Peliti1986}, branching random walks \cite{Cardy1998}, percolation \cite{Janssen2005}, phylogenetics \cite{Jarvis2005}, algebraic probability \cite{Ohkubo2013}, knot theory \cite{Rohwer2015}, and age dependent population dynamics \cite{Greenman2017}, to name a few.

These works have all been concerned with bosonic forms of QFT, where there is no restriction in occupation number. It is natural to consider the same question in a fermionic sense, where there can be no more than one particle in a state. This can be adapted to population dynamics by modeling classical motions of particles on a lattice, where sites are restricted to single maximum occupancy. Such an approach has been successfully used to model a range of systems such as aggregation processes \cite{Sandow1993}, Ising models \cite{Schulz1996Ising}, and lattice diffusion \cite{Schulz1997}, for example.  Exclusive dynamics have also been achieved within a bosonic framework \cite{vanWijland2001}. Grassmannian path integral techniques can also be adapted to such systems \cite{Schulz2005}, \cite{Silva2008}. 

In addition to bosonic (unrestricted) and fermionic (single occupancy) statistics, QFT has been developed for states with limited occupation number (partial exclusion). This was first developed by Green \cite{Green1953} and has since been well characterized with the aid of generalized paragrassmannian variables \cite{Ohnuki1982}, although no fundamental particles of this nature have been observed to date, and path integral formulations for these methods are not widespread \cite{Polychronakos1996}, \cite{Greenberg2004}. The Doi framework using parafermi statistics for stochastic systems with partial exclusion has developed for cyclic chemical reactions \cite{Schulz1996b}, and for diffusion \cite{Schulz1996}, \cite{Schutz1994}, although path integral techniques have not previously been considered. We turn to this problem and address this deficit with the work presented.

We also mention that significant work in renormalization with Doi-Peliti techniques have also been developed \cite{Peliti1986}, \cite{Lee1994}, \cite{Lee1995}, \cite{Cardy1998}, \cite{Tauber2005}, \cite{Tauber2014}, although such methods are not explored in this work. A recent review of Doi-Peliti approaches can be found in \cite{Weber2017}.

The systems that we shall apply these methods to are partially excluded lattice diffusion \cite{Schutz1994}, \cite{Schulz1996}, \cite{Schulz1997}, where maximum particle numbers are fixed over a lattice of sites, and birth-death processes with a carrying capacity, where population size is limited over a single site. The latter are also known as stochastic logistic growth or Verhulst models \cite{Bharucha1960}, \cite{Feller1939}. These are characterized by birth and death rates $\beta_n$ and $\mu_n$ which depend upon population size $n$ in some capacity limiting fashion. A linear birth rate $\beta_n = p-n$, for example, reduces as the population capacity $p$ is approached. Such linear systems can be analyzed using classical techniques \cite{Kendall1948}, \cite{Takashima1956}. However, the per individual rate $\frac{\beta_n}{n}=\frac{p}{n}-1$ is not very natural. A birth rate $\beta_n = n(p-n)$ has a linear per individual birth rate, and approaches zero as full capacity is reached. Although more natural, the quadratic nature makes this difficult to analyze analytically \cite{Kendall1948}, \cite{Kendall1949}. A death rate $\mu_n = \mu n$ has a constant death rate per individual, and approaches zero as the population empties, so is reasonably natural and the approach we take, although quadratic death rates could similarly be considered. 

The work is organized as follows. Section \ref{Back} develops a generalized Fock system system suitable for stochastic systems with partial exclusion, explaining the different Fock spaces found in the literature \cite{Schulz1996}, \cite{Schulz1996b}, \cite{Schutz1994}. Section \ref{Para} describes how generalized paragrassmannian algebras can be used to construct coherent states. Section \ref{PI Const} develops a coherent state path integral representation, demonstrating that the non-commutative nature of paragrassmannian variables means Magnus expansions \cite{Magnus1954}, \cite{Blanes2009} are required to construct path integral actions. Section \ref{Apps} considers applications to birth-death processes and diffusion. Conclusions in Section \ref{Conc} complete the work.


\section{Fock Spaces}
\label{Back}

\subsection{General Structure}

We assume in all that follows that the maximum occupancy of any site is $p$. We also assume, until otherwise stated, that we are dealing with a single site, with occupancy $n$. We let $a$ and $a^\dag$ represent annihilation and creation operators for a single site. The Green parafermi relations then take the form \cite{Green1953}, \cite{Ohnuki1982}

\begin{equation}
[a,[a^\dag,a]]=2a.
\label{Green}
\end{equation}

When Green introduced parastatistics, he used what is now referred to as the Green representation. In this formulation we have $p$ distinct occupational `bins', the $i^{\mathrm{th}}$ associated with standard Pauli operators $a_i$ and $a_i^\dag$. These obey standard anti-commutation relations
\begin{equation}
\{a_i,a_i^\dag\} = 1, \hspace{7mm} \{a_i,a_i\}=\{a_i^\dag,a_i^\dag\}= 0.
\label{Comm}
\end{equation}
These operators commute for distinct $i,j$, so $[a_i,a_j^\dag]=0$, for example. One can then show that operator $a = \sum_i a_i$ satisfies the Green relation of Eq. \ref{Green}. 

Next we introduce states $\ket{n}$ with $n \in \{0,1,\dots,p\}$ such that
\begin{equation}
a^\dag\ket{n} = p_n\ket{n+1},
\hspace{7mm}
a\ket{n} = q_n\ket{n-1},
\end{equation}
where $p_n,q_n$ are normalization factors that will later be specified. Repeated application of these recurrences results in
\begin{equation}
\ket{n}=\frac{(a^\dag)^n}{\prod_{i=0}^{n-1}p_i}\ket{0},
\hspace{7mm}
a^n\ket{n}=\prod_{i=1}^nq_i\ket{0}.
\label{RECCC}
\end{equation}
Now, the commutation relations can be applied to show that $a^n(a^\dag)^n\ket{0} = (n!)^2{p \choose n}\ket{0}$. We thus find from Eq. \ref{RECCC} that $(n!)^2{p \choose n} = \prod_{i=1}^np_{i-1}q_i$, which results in the expression
\begin{equation}
p_{n-1}q_n = n(p-n+1).
\end{equation}

This offers a range of possibilities for normalization factors $p_n,q_n$, five obvious choices of which are described in Table \ref{TableFock}. For the fermionic case ($p=1$) these choices are identical, whereas for fully parafermionic systems ($p>1$) they differ. The third Fock space was used for diffusion and a three species chemical reaction model in \cite{Schulz1996}, \cite{Schulz1996b}. The second Fock space was used for diffusion in \cite{Schutz1994}. It can be seen from a comparison between \cite{Schulz1996} and \cite{Schutz1994} that the second Fock space is algebraically easier to deal with than the third.

\begin{table}
\begin{center}
 \begin{tabular}[c]{c l l l l}
 \hline
 \\[-0.9em]
 Fock Space No. & $a^\dag\ket{n}=p_n\ket{n+1}$ & $a\ket{n}=q_n\ket{n-1}$ & $\braket{s|a^r|\psi}$ & $\braket{n|n}$ \\ [0.5ex] 
 \hline\hline
 1 & $(p-n)\ket{n+1}$ & $n\ket{n-1}$ & 
 $\sum_n\psi_n(p-n+r)_r$ & ${p \choose n}^{-1}$ \\ 
 \hline
 2 & $(n+1)\ket{n+1}$ & $(p-n+1)\ket{n-1}$ & $\sum_n\psi_n(n)_r$ & ${p \choose n}$ \\
 \hline
 3 & $\sqrt{(n+1)(p-n)}\ket{n+1}$ & $\sqrt{n(p-n+1)}\ket{n-1}$ & $\sum_n\psi_n\sqrt{(n)_r(p-n+r)_r}$ & $1$ \\
 \hline
 4 & $\ket{n+1}$ & $n(p-n+1)\ket{n-1}$ & $\sum_n\psi_n$ & $(n!)^2{p \choose n}$ \\
 \hline
 5 & $(n+1)(p-n)\ket{n+1}$ & $\ket{n-1}$ & $\sum_n\psi_n(n)_r(p-n+r)_r$ & $(n!)^{-2}{p \choose n}^{-1}$\\
 \hline
\end{tabular}
\caption{Fock space alternatives satisfying $p_nq_n = n(p-n+1)$. Terms $(n)_r = n(n-1)\dots(n-r+1)$ are Pochhammer symbols.}
\label{TableFock}
\end{center}
\end{table}

Next we introduce the number operator
\begin{equation}
N = p-[a^\dag,a]=a^\dag\cdot a = \sum_i a_i^\dag a_i.
\end{equation}
For all the Fock spaces in Table \ref{TableFock}, the states only differ in magnitude and satisfy the same eigenstate equation $N\ket{n}=n\ket{n}$. The number operator is thus identical across all Fock spaces.

Using this formalism, we let $\ket{s} = \sum_{n=0}^p\ket{n}$ (for the second Fock space, this is equivalent to the standard expression $\ket{s}=e^{a^\dag}\ket{0}$), and we let $\psi_n$ represent the probability that the site is occupied by $n$ individuals. Then, for all Fock spaces, we represent the state of the system as
\begin{equation}
\ket{\psi} = \sum_{n=0}^p \psi_n \ket{n}\braket{n|n}^{-1}.
\end{equation}

With this formalism, we can recover statistical features of interest. For example, in all cases the probability of $n$-fold occupation is given by
\begin{equation}
\psi_n = \braket{n|\psi}.
\label{ProbEq}
\end{equation}
For the second Fock space, we find that the $r^\textrm{th}$ factorial moment is
\begin{equation}
\braket{(n)_r}_\psi(t) = \sum_n (n)_r\psi_n(t) = \braket{s|a^r|\psi(t)},
\label{MnEq}
\end{equation}
where $(n)_r = n(n-1)\dots(n-r+1)$ denotes the Pochhammer symbol. This is the standard form usually observed for moments using bosonic Doi-Peliti methods \cite{Doi1976}, \cite{Doi1976b}, \cite{Peliti1985}. Note that for the remaining Fock spaces, the moment equations will differ (see Table \ref{TableFock}). However, from now on, we shall just be using the algebraically more compact second Fock space. 


\subsection{Liouvillians}

To model the stochastic dynamics of interest, we convert the corresponding master equation into the following form, where $\mathcal{L}$ denotes a suitable \emph{Liouvillian} operator:
\begin{equation}
\frac{d\ket{\psi}}{dt} = \mathcal{L}\ket{\psi}.
\label{KetDyn}
\end{equation}
From this formalism dynamic equations of interest can be readily obtained. For example, utilizing Eq. \ref{ProbEq}, the master equation is recovered via
\begin{equation}
\frac{\partial \psi_n}{\partial t} = \braket{n|\mathcal{L}|\psi}.
\end{equation}
Similarly, from Eq. \ref{MnEq}, we find factorial moment dynamic equation
\begin{equation}
\frac{\partial \braket{(n)_r}_\psi}{\partial t} = \braket{s|a^r\mathcal{L}|\psi(t)} = \braket{s|[a^r,\mathcal{L}]|\psi(t)},
\label{FMDE}
\end{equation}
where probability conservation $\bra{s}\mathcal{L}=0$ has been used in the latter form. 

We note finally that Eq. \ref{KetDyn} has formal solution
\begin{equation}
\ket{\psi(t)} = e^{t\mathcal{L}}\ket{\psi(0)}.
\end{equation}
This form will later be used to construct path integral representations of factorial moments of interest, offering an alternative approach to solving the dynamic equation of Eq. \ref{FMDE}. Next, however, we consider the dynamic form for some applications of interest. 


\subsection{Applications}

The three applications we consider are a birth-death processes with linear rates, one with quadratic rates, and a lattice diffusion process.

Firstly, then, consider a birth-death process where the population birth rate is $\beta_n = \beta(p-n)$ and death rate is $\mu_n = \mu n$. This is perhaps the simplest model of a birth-death system with carrying capacity, where the population is restricted in size between $0$ and $p$. The corresponding birth-death master equation is
\begin{equation}
\frac{d \psi_n}{dt} = -\psi_n (\beta_n+\mu_n)+\psi_{n+1}\mu_{n+1}+\psi_{n-1}\beta_{n-1}.
\label{MasterEq}
\end{equation}
The death term converts into bra-ket formalism as follows:
\begin{equation}
\sum_n\psi_{n+1}\mu(n+1)\ket{n}{p \choose n} = \sum_n\psi_{n+1}\mu(p-n)\ket{n}{p \choose n+1} = \sum_n\psi_{n+1}\mu a\ket{n+1}{p \choose n+1}=\mu a\ket{\psi}. 
\end{equation}
We can similarly convert all terms in the master equation to get a Liouvillian operator
\begin{equation}
\mathcal{L} = \beta (a^\dag-\overline{N}) + \mu (a-N),
\label{LinLiouv}
\end{equation}
where $\overline{N}=p-N$. From the commutation relations we find $[a,\mathcal{L}]=-2\beta N-\nu a+\beta p$, where $\nu = \mu-\beta$ and so, using Eq. \ref{FMDE} and $\gamma = \mu + \beta$, we find mean occupancy satisfies
\begin{equation}
\frac{\partial \braket{n}_\psi}{\partial t} = -\gamma \braket{n}_\psi +\beta p. 
\end{equation}
If we assume an initial population of size $n$ this results in solution
\begin{equation}
\braket{n}_\psi = \frac{\beta p}{\gamma}(1-e^{-\gamma T})+ne^{-\gamma T}.
\label{LinMean}
\end{equation}

For a second application, we consider a birth-death model with quadratic birth rate $\beta_n = \beta n(p-n)$ and linear death rate $\mu_n = \mu n$. This results in the two following Liouvillians, both equally valid:
\begin{eqnarray}
\mathcal{L} & = & -\beta N(p-N)-\mu N + \beta a^\dag N + \mu a,
\label{Quad1}\\
\mathcal{L} & = & -\beta a^\dag a-\mu N + \beta a^\dag N + \mu a.
\label{Quad2}
\end{eqnarray}
Although distinct operators, the moment equation $\frac{\partial \braket{n}_\psi}{\partial t}=\braket{s|[a,\mathcal{L}]|\psi}$ in both cases gives:
\begin{equation}
\frac{\partial \braket{n}_\psi}{\partial t} = -\braket{n}_\psi(\mu-\beta(p-1))-\beta \braket{n^2}_\psi, 
\end{equation}
which implicates the second moment. We could similarly derive an equation for the second moment, which would implicate higher moments, and so on, making the system more awkward to solve than the linear process analyzed above. We later turn to path integrals to provide some insight.

Finally, we consider diffusion on a lattice, with particles transferring from site $i$ to neighbouring site $j$ at rate $\nu n_i(p-n_j)$, where $n_i$, $n_j$ are the associated occupation numbers. Then the Liouvillian takes the form \cite{Schutz1994}
\begin{equation}
\mathcal{L} = \nu\sum_{i,j}\left(N_i\overline{N}_j-a_ia_j^\dag\right),
\label{DiffLiouv}
\end{equation}
where ordered pairs $(i,j)$ denote neighboring lattice sites. Note the subscripts $i,j$ are now referring to sites rather than bin indices of Eq. \ref{Comm}. Using the third Fock space in Table \ref{TableFock} results in a somewhat more complicated Liouvillian \cite{Schulz1996}. One can use the commutation relations to derive dynamic equations for moments using equations such as Eq. \ref{FMDE}. Equations for the first two moments are given below (the mean was observed in \cite{Schulz1996}), both taking the following discretized forms of diffusion, where $i(j)$ and $i(k)$ index neighbors of sites $j$ and $k$, respectively:
\begin{eqnarray}
\frac{\partial\braket{n_k}_\psi}{\partial t} & = & \nu p\sum_{i(k)}(\braket{n_j}_\psi-\braket{n_i}_\psi),\nonumber\\
\frac{\partial\braket{n_jn_k}_\psi}{\partial t} & = & \nu p\sum_{i(j)}\braket{(n_i-n_j)n_k}_\psi+\nu p\sum_{i(k)}\braket{n_j(n_i-n_k)}_\psi
+\left\{
\begin{array}{lr}
    -2\nu\sum_{i(k)}\braket{n_in_k}_\psi, & j=k,\\
    -2\nu\braket{n_jn_k}_\psi-\nu p\braket{n_j+n_k}_\psi, & |j-k|=1,\\
    0, & |j-k|>1.
  \end{array}
\right.
\end{eqnarray}

We are thus able to obtain dynamic equations for our examples using the Fock formalism directly. Solutions from these equations can then be sought. We now turn to alternative approaches afforded by path integral construction. In order to do this we first need the machinery of paragrassmannian algebras.


\section{Paragrassmannian Algebra}
\label{Para}

Next we introduce generalized paragrassmannian algebras that are required for coherent state path integral construction. Paragrassmannian vectors $\xi$ with components $\xi_i$, $i\in\{1,2,\dots,p\}$ are defined by \cite{Ohnuki1982}
\begin{equation}
\xi_i\xi_j = \eta_{i,j}\xi_j\xi_i,
\end{equation}
where the \emph{signature} $\eta$ satisfies $\eta_{i,i}=-1$ and $\eta_{i,j}=\eta_{j,i} \in \{-1,+1\}$. For our applications, we shall be interested in the parafermionic case, where $\eta_{i,j}=+1$ ($i \ne j$). The same commutation relations apply if one of the grassmannian variables, say $\xi_i$, is replaced with operator $a_i$ or $a_i^\dag$ (e.g. $\xi_i a_i^\dag = - a_i^\dag\xi_i$).

We introduce coherent states as:

\begin{equation}
\ket{\xi}  = e^{\xi \cdot a^\dag}\ket{\phi},
\hspace{7mm}
\bra{\xi^*}  = \bra{\phi}e^{a\cdot \xi^*},\\
\end{equation}
where we utilize dot product representations such as $\xi.a^\dag=\sum_{i=1}^p\xi_ia_i^\dag$. One can then use the commutation relations to establish the following eigenfunction and normalization properties:
\begin{equation}
\bra{\xi^*}a_i^\dag = \bra{\xi^*}\xi_i^*,
\hspace{7mm}
a_i\ket{\xi} = \xi_i\ket{\xi},
\hspace{7mm}
\braket{\xi^*|\xi} = e^{\xi^*\cdot\xi}.
\label{coh_norm}
\end{equation}

A coherent state path integral requires a resolution of the identity between time slices. Integration with respect to paragrassmannians is thus required, which is identical to grassmannian integration for each component:
\begin{equation}
\int \dd\xi_i\hh 1 = 0, \hspace{7mm} \int \dd\xi_i\hh \xi_i = 1.
\end{equation}
Note that integration acts from the left. For example, $\int \dd\xi_i\hh \zeta_i\xi_i =-\int \dd\xi_i\hh \xi_i\zeta_i = -\zeta_i$, where $\zeta_i$ is a distinct paragrassmannian. Now, from the definition of integration and coherent states, one can obtain the following resolution of identity (see appendices of \cite{Ohnuki1982} for a derivation):

\begin{equation}
I = \iint \dd\xi^* \dd\xi \hh e^{-\xi^*\cdot\xi} \ket{\xi}\bra{\xi^*},
\label{ResId}
\end{equation}
where $\dd\xi = \prod_{i=1}^p\dd\xi_i$ and $\dd\xi^* = \prod_{i=1}^p\dd\xi_i^*$. Note that the commutator relations $[\xi_i,\xi_j]=[\xi^*_i,\xi^*_j]=0$ for $i \ne j$ means the order of integration within each product is not important. However, $\dd\xi^*$ and $\dd\xi$ do not commute and their order matters.


\section{Path Integral Construction}
\label{PI Const}

We next use the paragrassmannians to construct a path integral, assuming the second Fock space of Table \ref{TableFock} in all that follows. The $r^\textrm{th}$ factorial moment, with resolutions of the identity, can be written as the following, where $z_0$ and $z_T$ are initial and final paragrassmannian vectors:
\begin{equation}
\braket{(n)_r}(T) = \sum_m (m)_r\psi_m(T) = \braket{s|a^re^{T\mathcal{L}}|\psi(0)} = \iint \dd z_T^*\dd z_T\dd z_0^*\dd z_0e^{-z_T^*\cdot z_T-z_0^*\cdot z_0}
\bra{s}a^r\ket{z_T}\bra{z_T^*}e^{T\mathcal{L}}\ket{z_0}\braket{z_0^*|\psi(0)}.
\end{equation}

This gives us an initial term $\braket{z_0^*|\psi(0)}$, a final term $\bra{s}a^r\ket{z_T}$, and a time evolution factor $\bra{z_T^*}e^{T\mathcal{L}}\ket{z_0}$ to calculate.

For the initial state we use $\ket{\psi(0)}=\ket{n}{p \choose n}^{-1}$ to represent an initial population size of $n$. Using the commutation relations, this gives us an initial term of the form $\braket{z_0^*|\psi(0)} = {p \choose n}^{-1}S_n(z_0^*)$, where $S_n(z_0^*)$ is composed of the sum of products of terms in all subsets of size $n$ from the set $\{z_1^*,\dots,z_p^*\}$ of components of paragrassmannian vector $z_0^*$. For example, with $p=4$, $S_2(z^*) = z_1^*z_2^*+z_1^*z_3^*+z_1^*z_4^*+ z_2^*z_3^*+z_2^*z_4^*+z_3^*z_4^*$. For the final term we find $\bra{s}a^r\ket{z_T} = z_T^re^{z_T}$, where $z_T$ is shorthand for $\sum_i(z_T)_i$.

To evaluate the time evolution factor, we first assume the Liouvillian operator $\mathcal{L}(a^\dag, a)$ has been put into normal form via the commutation relations, with creation operators left of annihilation operators. Complications arising from the general situation are described in \cite{Ohnuki1982}, however, for the models we consider, this is straightforwardly done. From this we can attempt to construct a coherent state path integral in the usual fashion. There is, however, a significant adjustment that we need to be aware of that arises due to the non-commutative nature of paragrassmannian variables. Now, we write the time evolution factor as a product of time slices with the aid of Eq. \ref{ResId} to give:
\begin{equation}
\braket{z_T|e^{T\mathcal{L}(a^\dag,a)}|z_0} =  \lim_{\epsilon\rightarrow 0}\prod_{\alpha=1}^N \iint \dd z_\alpha^* \dd z_\alpha \hh e^{-z_\alpha^*\cdot z_\alpha} \braket{z_\alpha^*|e^{\epsilon \mathcal{L}(a^{\dag},a)}|z_{\alpha-1}} 
 = \iint \mathcal{D}z^*\mathcal{D}z\hh e^{-\int_0^T \ddd t\hh \left(z^*(t)\cdot \frac{\partial z(t)}{\partial t}\right)-z^*(0)\cdot z(0)}\Pi(T),
\label{ProdDecomp}
\end{equation}
where $T=N\epsilon$, $z_\alpha$ represent paragrassmannian vectors for each $\alpha$, $z(t)$ is a time dependent paragrassmannian vector, and
\begin{equation}
\Pi(T) = \lim_{\epsilon\rightarrow 0}\prod_{\alpha=1}^{N-1} 
e^{\epsilon\mathcal{L}(z_\alpha^*,z_{\alpha-1})}.
\end{equation}

Note that terms such as $z_\alpha^*\cdot z_{\alpha-1}$ commute with all paragrassmannian variables and have been collected into a single exponential $\exp\left(\int_0^T\left(z^*(t).\frac{\partial z(t)}{\partial t}\right)-z^*(0)\cdot z(0)\right)$. It is standard methodology to try the same with terms $e^{\epsilon\mathcal{L}(z_\alpha^*,z_{\alpha-1})}$.  For some stochastic models of interest, however, these terms do not commute (see next section) and the continuous Baker-Campbell-Hausdorff theorem is required \cite{Blanes2009}. This results in the following time ordered Dyson series expansion, where $X(t) = \mathcal{L}(z^*(t),z(t))$, $\tau$ is the time ordering operator, and $\Delta_n$ is the triangular region $T>t_1>\dots >t_n>0$:
\begin{equation}
\Pi(T) = \tau\left\{\exp\int_0^T X(t)\hh\dd t\right\}
= 1 + \int_0^T  X(t_1)\hh\dd t_1 + \int_{\Delta_2} X(t_1)X(t_2)\hh\dd t_1\dd t_2+ \dots + \int_{\Delta_n} X(t_1)\dots X(t_n)\hh\dd{\bf t}_n+\cdots.
\label{Pert}
\end{equation}

Now, we are using paragrassmannian variables, thus to write the series $\Pi(T) = e^{\Omega(T)}$ as an exponential and construct an action requires a Magnus series \cite{Magnus1954}. There are many different formulations for these series \cite{Blanes2009}, the most explicit given by Saenza and Suarez \cite{Saenz2002}:
\begin{equation}
\Omega(T) = \int_0^T X(t_1)\hh\dd t_1 + \frac{1}{2}\int_{T>t_1>t_2>0}[X(t_1),X(t_2)]\hh\dd t_1\dd t_2+\sum_{n \ge 3}\frac{1}{n}\int_{\Delta_n}L_n[\dots[X(t_1),X(t_2)],\dots,X(t_n)]\hh\dd{\bf t}_n,
\end{equation}
where, given standard step function $\theta(t)$,
\begin{equation}
L_n = \sum_{i=1}^n\frac{(-1)^{i+1}}{i}\sum_{0<j_1<j_2\cdots<j_{n-i}<n}
\prod_{m=1}^{n-i}\theta(t_{j_m}-t_{j_m+1}).
\end{equation}
In the case that $X(t)$ is commutative (e.g. for bosonic systems) $\Omega(T)=\exp\int_0^TX(t)\hh\dd t$, resulting in the form usually observed in path integral actions \cite{Peliti1985}. Although the case of commutative $[X(t),X(s)]$ is tractable (see below), the general case results in a complicated action involving an infinite number of terms.

We next consider these path integrals for specific examples.


\section{Applications}
\label{Apps}


\subsection{Linear Birth and Death Process with Carrying Capacity}

Consider next the linear birth-death model with the Liouvillian given in Eq. \ref{LinLiouv}. We are interested in a path integral formulation for the $r^{\textrm{th}}$ factorial moment, $\braket{(n)_r}$, which should be equivalent to the expression given in Eq. \ref{LinMean} for $r=1$.

Now, for paragrassmannian variables $x^*$, $x$, $y^*$ and $y$, the commutator
\begin{equation}
[\mathcal{L}(x^*,x),\mathcal{L}(y^*,y)] = 2(\mu x + \beta x^*)\cdot(\mu x + \beta x^*),
\end{equation}
is commutative, meaning the Magnus expansion $\Omega(t)$ contains two terms and we find, using $\nu = \mu-\beta$, and shorthand notation $z_t=z(t)$,
\begin{eqnarray}
\braket{(n)_r}(T) & = & {p \choose n}^{-1}e^{-p\beta T} \iint \mathcal{D}z^*\mathcal{D}z \hh z_T^re^{z_T}\exp\left\{-\int_0^T \ddd t\hh (z_t^*\cdot(\partial_t z_t+\nu z_t))- z^*_0\cdot z_0\right\}\cdot
\label{LBDPI}\\
&& \hspace{30mm} \exp\left\{\int_0^T\ddd t\hh(\beta z_t^*+\mu z_t)+\int_0^T\ddd t\int_0^t\ddd s\hh(\beta z_t^*+\mu z_t)\cdot(\beta z_s^*+\mu z_s)\right\}
S_n(z^*_0).
\nonumber
\end{eqnarray}

Note that terms such as $z_t$ in dot products represent paragrassmannian vectors, otherwise it is shorthand for $\sum_i(z_t)_i$. Now, the first and third exponentials in the path integral contain non-commutative terms and the Baker-Cambell-Hausdorff theorem would be required to construct a single action for this path integral. However, we shall treat the first and third terms perturbatively. To do this we require a generating functional of the following form, rearranged using standard completing the square techniques \cite{Mandl2010}:
\begin{equation}
Z(\eta^*,\eta) = \iint \mathcal{D}z^*\mathcal{D}z e^{[z^*Dz]+[z^*\eta]+[\eta^*z]} = e^{-[\eta^*D^-\eta]}\iint \mathcal{D}z^*\mathcal{D}z e^{[z^*Dz]} = e^{-[\eta^*D^-\eta]}Z(0,0).
\end{equation}
Here we have used the shorthand notation $[z^*D^+z]=\int_0^T\int_0^T\dd t\dd s\hh z^*(t)\cdot D^+(t,s)z(s)$ and $[z^*\eta] = \int_0^T\dd t\hh z^*(t)\cdot \eta(t)$, where $D_{i,j}^+(s,t) = \delta_{i,j}\delta(t-s)(\partial_t+v+\delta(t))$ and the inverse $D_{i,j}^{-}(t,s)=\delta_{i,j}\theta(t-s)e^{-\nu(t-s)}$ satisfy equations $\sum_{j=1}^p\int_0^T \dd u\hh D_{i,j}^{\pm}(t,u)D^{\mp}_{j,k}(u,s)=\delta_{i,k}\delta(t-s)$, and $\eta^*$, $\eta$ are indeterminant paragrassmannian vectors.

To simplify the expansion, we make the following change of variables:
\begin{equation}
\left.
\begin{array}{l}
y = \alpha^{-1}(\beta z^* + \mu z)\\
y^* = \alpha^{-1}(\beta z^* - \mu z)\\
\end{array}
\right\}
\iff
\left\{
\begin{array}{l}
z = \frac{\alpha}{2\mu}(y-y^*),\\
z^* = \frac{\alpha}{2\beta}(y+y^*).\\
\end{array}
\right.
\label{Sub}
\end{equation}
The factor $\alpha = \sqrt{2\beta\mu}$ ensures the transformation has unit Jacobian, meaning that the path integral measure $\iint \mathcal{D}z^*\mathcal{D}z = \iint \mathcal{D}y^*\mathcal{D}y$ is preserved. Thus we get:
\begin{equation}
\braket{(n)_r} = {p \choose n}^{-1}e^{-p\beta T} \iint \mathcal{D}y^*\mathcal{D}y\hh z_T^re^{z_T}e^{[y^*Ey]}e^{\alpha\int_0^T\ddd t\hh y_t+\alpha^2\int_0^T\int_0^t\ddd t\ddd s\hh y_t\cdot y_s}S_n(z_0^*),
\end{equation}
where $[y^*Ey]$ represents $[z^*Dz]$ following substitution with Eq. \ref{Sub}. We also find that with the transformation
\begin{equation}
\left.
\begin{array}{l}
\zeta = \alpha^{-1}(\beta \eta^* + \mu \eta)\\
\zeta^* = \alpha^{-1}(\beta \eta^* - \mu \eta)\\
\end{array}
\right\}
\iff
\left\{
\begin{array}{l}
\eta = \frac{\alpha}{2\mu}(\zeta-\zeta^*),\\
\eta^* = \frac{\alpha}{2\beta}(\zeta+\zeta^*),\\
\end{array}
\right.
\end{equation}
the generating functional can be written as
\begin{equation}
Z(\eta^*,\eta) = \hat{Z}(\zeta^*,\zeta) = Z(0,0)\exp\left\{\frac{1}{2}\int_0^T\int_0^T\dd t\dd s\hh(\zeta(t)+\zeta^*(t))\cdot D^-(t,s)(\zeta(s)-\zeta^*(s))\right\},
\end{equation}
which gives rise to four classes of propagator ($\delta\zeta$ is shorthand for functional derivative $\frac{\partial}{\partial \zeta}$)
\begin{eqnarray}
G_{yy}^{i,j}(t,s) & = & \iint\mathcal{D}y^*\mathcal{D}y\hh y_i(t)y_j(s)e^{[y^*Ey]} = \delta\zeta^*_i(t)\delta\zeta^*_j(s)\hat{Z}(\zeta^*,\zeta)|_{\zeta\equiv\zeta^*\equiv 0} = \frac{1}{2}D_{i,j}^-(t,s),\nonumber\\
G_{yy^*}^{i,j}(t,s) & = & \iint\mathcal{D}y^*\mathcal{D}y\hh y_i(t)y_j^*(s)e^{[y^*Ey]} = -\delta\zeta_i^*(t)\delta\zeta_j(s)\hat{Z}(\zeta^*,\zeta)|_{\zeta\equiv\zeta^*\equiv 0} = \frac{1}{2}D_{i,j}^-(t,s),\nonumber\\
G_{y^*y}^{i,j}(t,s) & = & \iint\mathcal{D}y^*\mathcal{D}y\hh y_i^*(t)y_j(s)e^{[y^*Ey]} = -\delta\zeta_i(t)\delta\zeta_j^*(s)\hat{Z}(\zeta^*,\zeta)|_{\zeta\equiv\zeta^*\equiv 0} = -\frac{1}{2}D_{i,j}^-(t,s),\nonumber\\
G_{y^*y^*}^{i,j}(t,s) & = & \iint\mathcal{D}y^*\mathcal{D}y\hh y_i^*(t)y_j^*(s)e^{[y^*Ey]} = \delta\zeta_i(t)\delta\zeta_j(s)\hat{Z}(\zeta^*,\zeta)|_{\zeta\equiv\zeta^*\equiv 0} = -\frac{1}{2}D_{i,j}^-(t,s).
\label{Prop}
\end{eqnarray}

Next consider the perturbative expansion. All the internal nodes of the associated Feynman diagram arise from expansion of the third exponential of Eq. \ref{LBDPI} into a Dyson series of Eq. \ref{Pert}, which has a general term $\alpha^n\int_{\Delta_n}y(t_1)\dots y(t_n) d{\bf t}_n$ to consider. Note, therefore, that the internal nodes are just associated with variable $y$ (rather than $y^*$).

Now, the initiating terms from $S_n(z_0^*)$ arise from components of the paragrassmannian vector $z_0^* = \frac{\alpha}{2\beta}(y_0+y_0^*)$. We see from above that propagators with initiating nodes $y_0$ or $y_0^*$ will give the same value (e.g $G_{yy^*}^{i,j}(t,s) =G_{yy}^{i,j}(t,s)$). We can thus represent this as a single $y$ node with factor $\frac{\alpha}{\beta}$. Similarly, we can replace any terminating nodes arising from $z_Te^{z_T}$ by $y$ nodes with factors $\frac{\alpha}{\mu}$. All nodes are thus of type $y$ and $G_{yy}^{i,j}(t,s)$ is the only propagator we need to consider.

From Eq. \ref{Prop} we note that propagators between node pairs in bins $i$, $j$ are only non-zero if $i=j$, and so are represented as horizontal intervals $([t_1,t_2],i)$ sitting within $[0,T]\times\{1,\dots,p\}$. Furthermore, the anti-commutative nature of paragrassmanian components with the same bin number $i$ means that time intervals of propagators do not overlap for each $i$. 

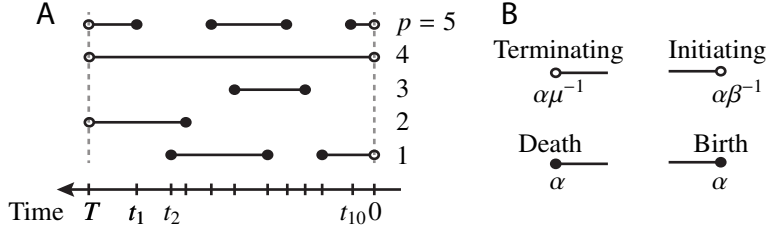
\begin{figure}[!t]
   \centering
   \setlength{\unitlength}{0.1\textwidth}
   \begin{picture}(6,2)
     \put(0,0){\includegraphics[width=10cm]{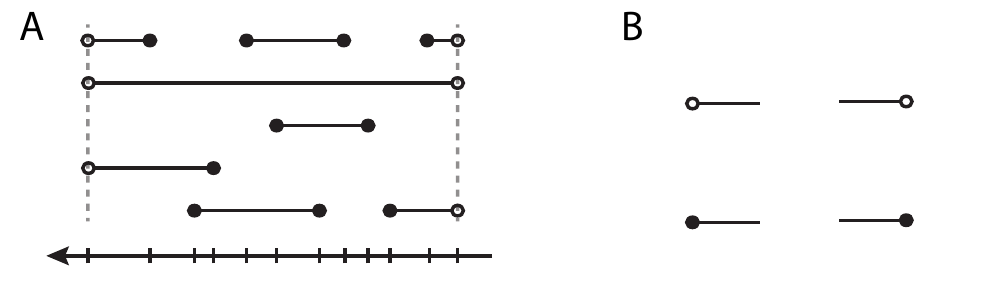}}
     \put(-0.1,0){Time}
     \put(2.78,0){$0$}\put(2.55,0){$t_{10}$}\put(1.14,0){$t_2$}
     \put(0.86,0){$t_1$}\put(0.50,0){$T$}
     \put(0.86,0){$t_1$}\put(0.50,0){$T$}
     \put(3.00,0.45){$1$}
     \put(3.00,0.72){$2$}
     \put(3.00,0.99){$3$}
     \put(3.00,1.26){$4$}
     \put(3.00,1.53){$p=5$}
     \put(3.78,1.3){Terminating}\put(3.98,0.55){Death}
     \put(4.10,0.95){$\alpha\mu^{-1}$}\put(5.53,0.95){$\alpha\beta^{-1}$}
     \put(5.18,1.3){Initiating}\put(5.38,0.55){Birth}
     \put(4.23,0.25){$\alpha$}\put(5.53,0.25){$\alpha$}
   \end{picture}
   \caption{Feynman diagram for linear birth death model. A) Sample  diagram. B) Diagram nodes.}
   \label{LinBDFig}
\end{figure}

The collections of intervals for each bin $i$ group into four classes depending upon whether they contain initiating or terminating nodes. Consider then the Laplace transform $\hat{I}_{01}(s)$ of the sum of all propagator products, within a single bin, of the class with one terminating node and no initiating node (e.g. the propagators in Fig. \ref{LinBDFig}A are the contribution from one diagram for $i=2$). This is a sum of convolutions of propagators $e^{-\nu t}$ (segments) and the value $1$ (gaps), giving a Laplace transform and inverse of the following form, where $\gamma = \beta+\mu$, and $k$ counts segment/gap pairs:
\begin{equation}
\hat{I}_{01}(s) = \frac{1}{\mu}\sum_{k=1}^\infty\frac{(\beta\mu)^k}{s^k(s+\nu)^k} =\frac{\beta}{(s-\beta)(s+\mu)},
\hspace{7mm}
I_{01}(T) = \frac{\beta}{\gamma}(e^{\beta T}-e^{-\mu T}).
\end{equation}

The other three cases are found similarly, giving $I_{10}(T) = \frac{\mu}{\beta}I_{01}(T)$, $I_{00}(T) = \frac{\mu}{\gamma}e^{\beta T}+\frac{\beta}{\gamma}e^{-\mu T}$ and $I_{11}(T) = \frac{\beta}{\gamma}e^{\beta T}+\frac{\mu}{\gamma}e^{-\mu T}$. Now the sum of terms across all Feynman diagrams for each bin $i \in \{1,2,\dots,p\}$ contains one of these four options as a factor in a product of $p$ terms. Given $n$ initial nodes, which are counted by copies of $I_{10}(T)$ and $I_{11}(T)$, we sum over the possibilities to give the $r^\textrm{th}$ factorial moment as a hypergeometric sum
\begin{equation}
\braket{(n)_r}_\psi = Z(0,0)e^{-p\beta t}
\sum_{a=0}^{p-n}\sum_{b=0}^n(a+b)^r{p-n \choose a}{n \choose b}{I_{01}}^a{I_{00}}^{p-n-a}{I_{11}}^b{I_{10}}^{n-b}.
\end{equation}
If we calculate this for the $0^\textrm{th}$ mean ($\braket{1}_\psi=1$) to find $Z(0,0)$, the mean ($r=1$) can then be found, which reduces to the formulation given in Eq. \ref{LinMean} after some algebra, as one might expect. Higher moments can be found similarly.


\subsection{Quadratic Birth Death Process}

Next consider the quadratic birth-death process, where we found associated Liouvillians in Eq. \ref{Quad1}, \ref{Quad2}. To construct a path integral approach, we note that $[\mathcal{L}(x^*,x),\mathcal{L}(y^*,y)]=2(\beta(x^*\cdot x)x^*+\mu x)\cdot(\beta(y^*\cdot y)y^*+\mu y)$, using the Liouvillian of Eq. \ref{Quad1}. This is a commutative object meaning we have a finite Magnus expansion with two terms. Note that this is not the case with the Liouvillian of Eq. \ref{Quad2}, where the innocuous looking term $a^\dag a$ results in a complicated Magnus expansion. Using the former results in a factorial moment path integral of the following form, where $\nu = \mu+\beta(p-1)$,
\begin{eqnarray}
\braket{(n)_r}(T) & = & {p \choose n}^{-1} \iint \mathcal{D}z^*\mathcal{D}z\hh z_T^re^{z_T}\exp\left\{-\int_0^T \ddd t\hh(z_t^*\cdot(\partial_t z_t+\nu z_t))- z^*_0\cdot z_0+\beta\int_0^T\ddd t\hh(z_t^*\cdot z_t)^2\right\}
\cdot\label{QuadBDPI}\\
&&\hspace{10mm}
\exp\left\{\int_0^T\ddd t(\beta (z_t^*\cdot z_t)z_t^*+\mu z_t)+\int_0^T\ddd t\int_0^t\ddd s\hh(\beta(z_t^*\cdot z_t) z_t^*+\mu z_t)\cdot(\beta (z_s^*\cdot z_s) z_s^*+\mu z_s)\right\}\nonumber
S_n(z^*_0).
\end{eqnarray}

\begin{wrapfigure}{t}{0.36\textwidth}
\centering
\setlength{\unitlength}{0.1\textwidth}
\begin{picture}(5,4.9)
\put(0,0){\includegraphics[width=5cm]{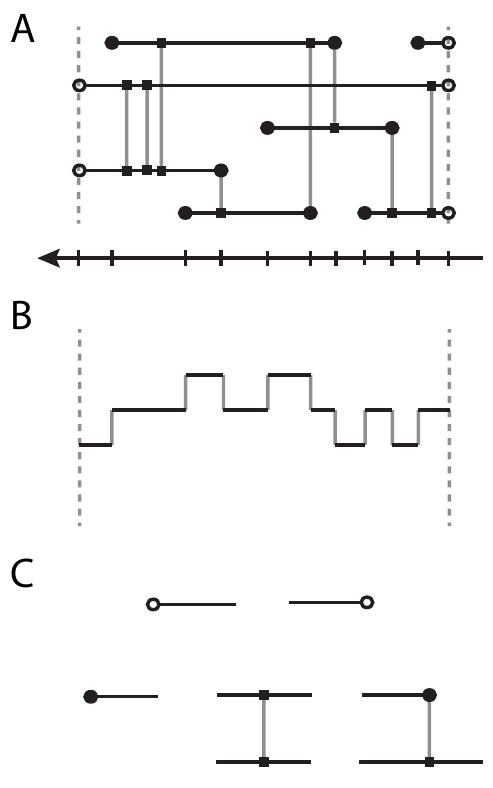}}
\put(-0.29,3.2){Time}
\put(2.72,3.00){$0$}\put(2.55,3.00){$t_{8}$}\put(1.30,3.00){$t_3$}
\put(1.10,3.00){$t_2$}\put(0.65,3.00){$t_1$}\put(0.45,3.00){$T$}
\put(3.00,3.45){$1$}\put(3.00,3.72){$2$}
\put(3.00,3.99){$3$}\put(3.00,4.26){$4$}
\put(3.00,4.53){$p=5$}
\put(3.00,1.83){$1$}\put(3.00,2.05){$2$}
\put(3.00,2.27){$3$}\put(3.00,2.49){$4$}
\put(3.00,2.71){$5$}\put(3.00,1.61){$0$}
\put(0.48,1.3){Terminating}\put(0.30,0.70){Death}
\put(1.88,1.3){Initiating}\put(2.45,0.7){Birth}
\put(0.37,0.48){$\mu$}\put(0.60,0.42){$z_i$}
\put(2.73,0.48){$\beta$}\put(2.47,0.40){$z^*_j$}
\put(2.47,0.00){$z^*_i$}\put(2.70,0.02){$z_i$}
\put(1.32,0.7){Neutral}\put(1.90,0.30){$\beta$}
\put(1.45,0.00){$z^*_i$}\put(1.68,0.02){$z_i$}
\put(1.45,0.40){$z^*_j$}\put(1.68,0.42){$z_j$}
\put(0.98,0.96){$(z^*_T)_i$}\put(1.95,0.96){$(z_0)_i$}
\end{picture}
\caption{Feynman diagram for quadratic birth death model. A) Sample diagram. B) Total propagator profile corresponding to A). C) Diagram nodes. }
\label{QuadBDFig}
\end{wrapfigure}
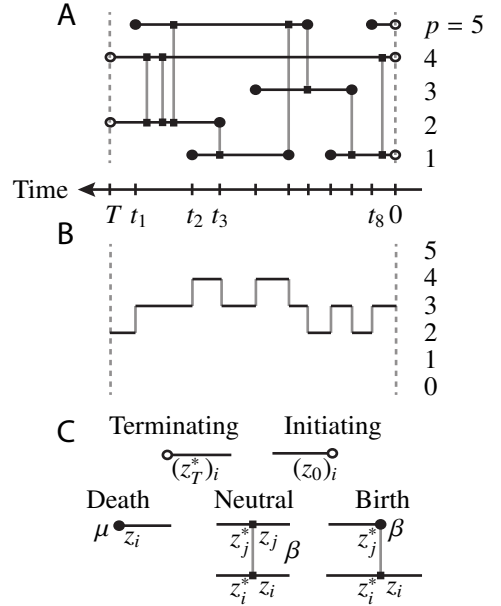 

We treat this with perturbative expansion, much the same as the previous example. The main difference is that instead of a birth term $\beta z^*$, we have a birth term $\beta (z^*\cdot z)z^*$, and we have a new term of the form $\beta(z^*\cdot z)^2$ to consider. We also have the same death term $\mu z$. This gives three classes of internal node in corresponding Feynman diagrams, as exemplified in Fig. \ref{QuadBDFig}C. 

The death term $\mu z=\mu\sum_iz_i$ results in a simple terminating node, where a node in bin $i$ corresponds to term $\mu z_i$. The term $\beta(z^*\cdot z)^2$ is composed of terms $2\beta z_i^*z_iz_j^*z_j$ with distinct $i,j$. This results in the vertical connections (labeled neutral in Fig. \ref{QuadBDFig}C) given in the sample diagram Fig. \ref{QuadBDFig}A. The birth term consists of terms of the form $\beta z_i^*z_iz_j^*$ with distinct $i,j$, represented as birth nodes with connections from an existing propagator in bin $i$ to a new propagator in bin $j$ (e.g time $t_3$ in Fig. \ref{QuadBDFig}A has a birth term for $i=1,j=2$). The propagator takes the same form $e^{-\nu t}$ as the previous section, albeit with a different $\nu$. Now the technique utilized in the previous section summed the Laplace transform of propagators for each bin $i \in \{1,\dots,p\}$ and then amalgamated the results. Now, however, the links between segments of differing bins arising from birth and neutral terms make this impossible and we need to change the approach, and instead will sum all diagrams collectively.

In Fig. \ref{QuadBDFig}B we see a diagram counting the number of propagators bridging any point in time. Note that in any time interval, any number of neutral connectors can be placed between two propagators without affecting the propagator count in that interval. For a time interval of length $t$ such that the propagator count $m$ is constant for each point of time (a flat portion of the path in Fig. \ref{QuadBDFig}B), we sum the Laplace transform of the product of propagators $e^{-\nu m t}$ of all these possibilities to give the following, where $\ell$ indexes the number of neutral connections,

\begin{equation}
\sum_{\ell=0}^\infty{m \choose 2}^\ell\frac{(2\beta)^\ell}{(s+m\nu)^{\ell+1}}=\frac{1}{s+m\nu-\beta m(m-1)} = \frac{1}{s+\mu_m+\beta_m}.
\label{Lapp}
\end{equation}

Note the following characteristics of Fig. \ref{QuadBDFig}B. The count profile can be represented as a path that can move one position up or down at a time corresponding to internal birth or death nodes, starting at the initial number of occupied bins ($n=3$ in Fig. \ref{QuadBDFig}B). The final value indicates the number of terminating nodes in the diagram ($n=2$ in our example). There are ${p \choose n}$ possible ways of selecting the $n$ initial nodes, which cancels the factor ${p \choose n}^{-1}$ in Eq. \ref{QuadBDPI}. If the path is at height $m$ prior to an upward step representing term $\beta z_i^*z_iz_j^*$, there are $p-m$ choices for the internal birth node (i.e. selecting an unoccupied bin $j$), and $m$ choices for occupied bin $i$. With weight $\beta$ these choices are encapsulated by factor $\beta_m=\beta m(p-m)$. If the path is at height $m$ prior to a drop representing term $\mu z_i$, there are $m$ choices (i.e. select an occupied bin $i$ to terminate), each associated with death factor $\mu$. These choices are encapsulated by factor $\mu_m=\mu m$. Now each time segment of height $m$ has a Laplace transform $\frac{1}{s+\mu_m+\beta_m}$ corresponding to a time dependent term $e^{-(\mu_m+\beta_m)t}$. Then the Laplace transform for all diagrams corresponding to a particular path with $k$ steps is given by:
\begin{equation}
\mathcal{L}\left\{\int_{0=t_0<t_1<\dots<t_k=T}\dd{\bf t}_k\hh\prod_{i=1}^ke^{-(\mu_{m_i} +\beta_{m_i})(t_i-t_{i-1})}\right\}
= \prod_{i=1}^k\frac{1}{s+\mu_{m_i}+\beta_{m_i}}.
\end{equation}

Next we introduce $f_m^{(k)}$ as the sum of Laplace transforms for all paths ending at height $m$ in $k$ steps. Note that $k-1$ counts the number of internal birth-death (i.e. non-neutral) nodes at time points separating steps. The implicit initial height $n$ is fixed throughout the discussion. Then we can construct the following recurrence:
\begin{equation}
f_m^{(k+1)} = \frac{1}{s+\mu_m+\beta_m}\left[\beta_{m-1}f_{m-1}^{(k)} +\mu_{m+1}f_{m+1}^{(k)}\right].
\end{equation}
Next introduce $f_m = \sum_{k \ge 1}f_m^{(k)}$. Then summing over the recurrence gives
\begin{equation}
f_m = f_m^{(1)}+\frac{1}{s+\mu_m+\beta_m}\left[\beta_{m-1}f_{m-1}+\mu_{m+1}f_{m+1}\right]  .
\end{equation}
Noting the initial condition $f_m^{(1)}=\delta_{mn}\frac{1}{s+\mu_m+\beta_m}$, we can write this as:
\begin{equation}
(s+\mu_m+\beta_m)f_m = \delta_{m,n}+\beta_{m-1}f_{m-1}+\mu_{m+1}f_{m+1}.
\label{Recc}
\end{equation}

Now the $r^\textrm{th}$ factorial moment $\braket{(n)_r}(T)$ of the population size can be written as a path integral with terminating nodes described by the factor $z_T^re^{z_T}$. Then we can write this as an inverse Laplace transform:
\begin{equation}
\braket{(n)_r}(T) = Z(0,0)\mathcal{L}_T^{-1}\left\{\sum_m (m)_rf_m\right\}.
\end{equation}

For the case $r=0$ we sum Eq. \ref{Recc} over $m$ to find $\sum_mf_m = s^{-1}$ resulting in $Z(0,0)=1$. Now if we write Eq. \ref{Recc} in matrix form as $(sI-B){\bf f} = {\bf e}_n$ with tridiagonal matrix $B$ we can write the mean as follows, where $\lambda_i$ represent eigenvalues of matrix $B$:
\begin{equation}
\braket{n}_\psi(T) = \mathcal{L}_T^{-1}\left\{\sum_m m f_m\right\} = \mathcal{L}_T^{-1}\left\{\sum_m m\left\{\textrm{Adj}(sI-B)\right\}_{mn}|sI-B|^{-1}\right\} = \sum_i\frac{\sum_m m\left\{\textrm{Adj}(\lambda_iI-B)\right\}_{mn}}{\prod_{j \ne i}(\lambda_i-\lambda_j)}e^{\lambda_it}.
\label{QuadSol}
\end{equation}

Now if vector ${\bf \psi}$ represents the population probability distribution, where component $\psi_m(T)$ is the probability of population size $m$, the master equation is $\frac{\partial {\bf \psi}}{\partial t} = B{\bf \psi}$, which has a solution of the form ${\bf p} = e^{TB}{\bf e}_n$. Diagonalizing $B$ results in precisely the solution given in Eq. \ref{QuadSol}, and we find that the perturbative expansion is equivalent to diagonalization of the original system \cite{Gantmakher1998}.


\subsection{Diffusion on a Lattice}

We lastly point out that these path integral techniques can be extended to lattice methods. Factorial moment path integral formulation is similar to the previous example, so we just highlight the salient points. Firstly, from Eq. \ref{DiffLiouv} we separate the Liouvillian $\mathcal{L}=\mathcal{L}^{0}+\mathcal{L}^{1}$ into a quadratic part $\mathcal{L}^{0}=-p\nu\sum_{i,j}N_i$ which we shall deal with non-perturbatively, and $\mathcal{L}^{1} = \nu\sum_{i,j}(N_iN_j+a_ia_j^\dag)$, which we deal with perturbatively. 

We rewrite Eq. \ref{ProdDecomp} as follows, where $\mathcal{L}^{0/1}_\alpha=\mathcal{L}^{0/1}(z_\alpha^*,z_{\alpha-1})$, $X(t) = \mathcal{L}^1(z_t^*,z_t)$, $z_\alpha$ is a paragrassmannian vector over bins and sites ($\alpha$ indexes time slices), and $z_t$ is shorthand for $z(t)$:
\begin{eqnarray}
\braket{z_T|e^{T\mathcal{L}(a^\dag,a)}|z_0} & = &  \lim_{\epsilon\rightarrow 0}\prod_{\alpha=1}^N \iint \dd z_\alpha^* \dd z_\alpha\hh e^{-z_\alpha^*\cdot z_\alpha} \braket{z_\alpha^*|e^{\epsilon \mathcal{L}(a^{\dag},a)}|z_{\alpha-1}} 
= \iint \mathcal{D}z^* \mathcal{D}z\hh
\lim_{\epsilon\rightarrow 0}
\prod_{\alpha=1}^N\braket{z_\alpha^*|z_{\alpha-1}}
(1+\epsilon\mathcal{L}_\alpha^0+\epsilon\mathcal{L}_\alpha^1)\nonumber\\
& = & \iint \mathcal{D}z^* \mathcal{D}z
\lim_{\epsilon\rightarrow 0}
\prod_{\alpha=1}^N\braket{z_\alpha^*|z_{\alpha-1}}
(1+\epsilon\mathcal{L}_\alpha^0)(1+\epsilon\mathcal{L}_\alpha^1)
= \iint \mathcal{D}z^*\mathcal{D}ze^{-\int z_t^*\cdot\left[ \frac{\partial z_t}{\partial t}+pd\nu z_t\right]-z_0^*\cdot z_0}\Pi(T),
\end{eqnarray}
where $d$ represents the number of neighbours of each site, and
\begin{equation}
\Pi(T) = \lim_{\epsilon\rightarrow 0}\prod_{\alpha=1}^{N-1} 
e^{\epsilon\mathcal{L}^1(z_\alpha^*,z_{\alpha-1})}
=\tau\left\{\exp\int_0^TX(t)dt\right\}
= \sum_{n=0}^\infty \int_{\Delta_n}X(t_1)\dots X(t_n)d{\bf t}_n.
\end{equation}

This produces a path integral with form similar to the previous section, except propagators take the form $e^{\nu pdt}$, we have two types of Feynman diagram nodes corresponding to the two types of terms in $\mathcal{L}^1$, and now there are $p$ bins for each site of the lattice.

Now, each term arising from $\nu\sum_{i,j}N_i N_j$ takes the form $\nu z^*_{i,\kappa}z_{i,\kappa}z^*_{j,\ell}z_{j,\ell}$ linking a propagator with label $(i,\kappa)$ to $(j,\ell)$, where $i$ and $j$ and neighboring sites, and $\kappa$ and $\ell$ index bins. These are analogous to the vertical segments representing neutral terms in the previous section. If ${\bf n}$ is a vector indexing the (assumed finite) number of occupied bins across sites, the number of possible links is $\pi_{\bf n} = \frac{1}{2}\sum_{i,j}n_in_j$. Then, much like Eq. \ref{Lapp}, the Laplace transform of the sum of such terms across any time segment with total occupation number $n=\sum_in_i$ is
\begin{equation}
L(s) = \sum_{m=1}^\infty\frac{(\nu\pi_{\bf n})^{m-1}}{(s+n\nu pd)^m}=\frac{1}{s+n\nu pd-\nu\pi_{\bf n}}.
\end{equation}

Now we assume that ${\bf n}^0$ is the initial distribution across the lattice. We let $f_{\bf n}^{(k)}$ denote the sum of Laplace transformed diagrams representing $k-1$ non-neutral nodes arising from $\nu\sum_{i,j}a_ia_j^\dag$ that start with distribution ${\bf n}^0$ and end with distribution ${\bf n}$. Then if ${\bf n}^{+i,-j}$ is the vector ${\bf n}$ with the $i^\textrm{th}$($j^\textrm{th})$ component increased(decreased) by one unit, we have recurrence:  
\begin{equation}
f_{\bf n}^{(k+1)} = \frac{\nu}{s+n\nu pd-\nu\pi_{\bf n}}\sum_{i,j}
f_{{\bf n}^{+i,-j}}^{(k+1)}(n_i+1)(p-n_j+1).
\end{equation}
Then defining $f_{\bf n}=\sum_{k=1}^\infty f_{\bf n}^{(k)}$ and noting that $(s+n\nu pd-\nu\pi_{\bf n})f_{\bf n}^{(1)} = \delta_{{\bf n},{\bf n}^0}$
results in a recurrence, analogous to Eq. \ref{Recc},
\begin{equation}
(s+n\mu+\pi_{\bf n})f_{\bf n} + \sum_{i,j}\nu
f_{{\bf n}^{+i,-j}}(n_i+1)(p-n_j+1) = \delta_{{\bf n},{\bf n}^0}.
\end{equation}
This is a finite (albeit high dimensional) system of equations for a finite population, which can be solved exactly, meaning moments can then be obtained in much the same way as the previous section.


\section{Conclusions}
\label{Conc}

In this paper a generalized Fock space has been developed, explaining the different Fock spaces seen in the literature for stochastic systems with partial exclusivity. The machinery of generalized paragrassmannian algebras has also been utilized to construct path integral formulations for features of interest, a methodology not explored elsewhere. The non-commutative nature of paragrassmannians means that a Magnus expansion is required to construct a single action, which can be complex in nature, resulting in actions with an infinite number of terms. This adjustment also applies to fermionic path integrals, an issue not discussed in the literature \cite{Schulz2005}, \cite{Silva2008}, \cite{Tauber2014}.

These methods have been applied to birth-death processes with a carrying capacity, and diffusion on an occupation limited lattice, producing results consistent with those obtained via classical methods. Finding alternative perturbative and non-perturbative expansion schemes offering alternative solution formulations, and maker greater use of the Magnus expansion, remains a future direction of research. An obvious approach is to examine the effect of the Doi shift on expressions. Although the quantity $\hat{z}=z+1$ is not a paragrassmannian number, $\int dz (a+bz)$ is invariant to such a substitution and a Doi shift is valid. However, the fact that the path integrals are not easily expressed as a single action seems to limit the usefulness of such an approach, and other ideas are needed.

\newpage




\section*{References}
\bibliographystyle{elsarticle-num} 
\bibliography{refs_DoiPelitiExcl}






\end{document}